\documentclass[amsmath,amssymb,11pt,superscriptaddress,reprint, preprintnumbers, notitlepage,aps,prl,twocolumn, dvipsnames,table]{revtex4-1}
\pdfoutput=1 
\usepackage[utf8]{inputenc}
\usepackage[english]{babel}
\usepackage{amsmath}
\usepackage{graphicx}
\usepackage{dcolumn}
\usepackage{pbox}
\usepackage{amssymb}
\usepackage{epsfig}
\usepackage[dvipsnames]{xcolor}
\usepackage[normalem]{ulem}
\usepackage{slashed}
\usepackage{amssymb}
\usepackage{verbatim} 
\usepackage{ mathrsfs }
\usepackage{color}
\usepackage[font=small]{caption}
\usepackage[font=small]{subcaption}
\usepackage{url}
\definecolor{MyDarkBlue}{rgb}{0.1, 0.1, 0.8}
\definecolor{SBlue}{rgb}{0.2, 0.4, 0.7} 
\definecolor{MyLightBlue}{rgb}{0.22,0.51,0.9}
\definecolor{MyGreen}{rgb}{0.0, 0.5, 0.0}
\definecolor{BrickRed}{rgb}{0.8, 0.25, 0.33}
\RequirePackage{hyperref}
\hypersetup{colorlinks, citecolor=SBlue,linkcolor=MyDarkBlue, urlcolor=PineGreen}

\usepackage{amsmath}
\usepackage{amssymb}
\usepackage{array}
\usepackage{multirow}
\usepackage{booktabs}
\usepackage{xcolor}

\makeatletter

\makeatletter
\renewcommand\@makecaption[2]{%
  \par
  \vskip\abovecaptionskip
  \begingroup
  
   \small\rmfamily
    \begingroup
     \samepage
     \flushing
     \let\footnote\@footnotemark@gobble
     \@make@capt@title{#1}{#2}\par
    \endgroup
  \endgroup
  \vskip\belowcaptionskip
}
\makeatother

\DeclareUnicodeCharacter{2212}{-}
\begin{document}

\preprint{HRI-RECAPP-2025-09}
\preprint{MS-TP-25-19}

\title{Gravitational Wave Signature and the  Nature of \\  Neutrino Masses: Majorana, Dirac, or Pseudo-Dirac?}

\author{\bf Sudip Jana}
\email[E-mail:]{sudip.jana@okstate.edu}
\affiliation{Harish-Chandra Research Institute, A CI of Homi Bhabha National Institute,
Chhatnag Road, Jhunsi, Allahabad, 211019, India}
\author{\bf Sudip Manna}
\email[E-mail:]{sudipmanna@hri.res.in}
\affiliation{Harish-Chandra Research Institute, A CI of Homi Bhabha National Institute,
Chhatnag Road, Jhunsi, Allahabad, 211019, India}
\author{\bf Vishnu P.K.}
\email[E-mail:]{vishnu.pk@uni-muenster.de}
\affiliation{Institut für Theoretische Physik, Universität Münster, Wilhelm-Klemm-Straße 9, 48149 Münster,
Germany}
\begin{abstract} 
The fermionic nature of neutrinos and the origin of their tiny masses remain unresolved issues in particle physics, intrinsically connected to lepton number symmetry—conserved for Dirac, violated for Majorana, and effectively pseudo-Dirac when global symmetries invoked for conservation are broken by quantum gravity. We investigate whether distinctive gravitational-wave (GW) signatures can illuminate the nature of neutrino masses and their underlying symmetries, particularly in scenarios where Yukawa couplings are not unnaturally small. To this end, we consider the minimal $B-L$ gauge extension of the Standard Model, where quantum numbers of beyond-SM states determine the neutrino nature and the scale of spontaneous $B-L$ breaking governs mass generation. In this framework, we show that neutrinos yield characteristic GW spectra: Majorana neutrinos with high-scale breaking ($\sim 10^{14}$ GeV) produce local cosmic strings and a flat spectrum across broad frequencies, Dirac neutrinos with low-scale breaking ($\sim 10^{7}$ GeV) generate peaked spectra from first-order phase transitions, and pseudo-Dirac scenarios give kink-like features from domain wall annihilation.
\end{abstract}
\maketitle
\textbf{\emph{Introduction}.--}
The advent of gravitational-wave (GW) astronomy opened a new window in exploring the Universe, marked by a landmark observation of a binary neutron star merger \cite{LIGOScientific:2017vwq}, detected through gravitational waves and promptly followed by a short gamma-ray burst. This frontier advanced further with the detection of a GW event accompanied by neutrinos \cite{GCN38679}. Beyond confirming neutrino sources, this milestone raises a deeper question: can such signals shed light on the theory of neutrino oscillations—firm evidence of tiny but nonzero neutrino masses and, hence, of physics beyond the Standard Model? Despite worldwide efforts across energy, intensity, precision, and cosmic frontiers, the origin of neutrino mass remains unresolved. Against this backdrop, the prospect that the GW signal might provide insights into the origin of neutrino mass, or even reveal the underlying symmetry that governs it, would represent a profound breakthrough.

Another fundamental question, closely tied to the above discussion, concerns the nature of neutrinos: are they Dirac, Majorana, or Pseudo-Dirac particles, and how can these possibilities be distinguished? Neutrinoless double beta decay (\(0\nu\beta\beta\)) has long been regarded as the most promising probe of the Majorana hypothesis through lepton-number violation \cite{Schechter:1981bd}. While its unambiguous detection would indeed establish neutrinos as Majorana fermions, important caveats persist \cite{Graf:2023dzf}. The observed signal may be dominated by new-physics contributions unrelated to neutrino masses, while the direct contribution from the Majorana mass term itself could remain vanishingly small \cite{Duerr:2011zd}.  Various other approaches have also been explored to test the Dirac, Majorana, or Pseudo-Dirac nature of neutrinos in different contexts \cite{Balantekin:2018ukw, Hernandez-Molinero:2022zoo, Lunardini:2019zob,  Jana:2022tsa, Jana:2023ufy, Carloni:2025dhv, Rosen:1982pj, Rodejohann:2017vup, Berryman:2018qxn}.  

\begin{figure}[htb!]
    \centering
    \includegraphics[width=0.5\textwidth]{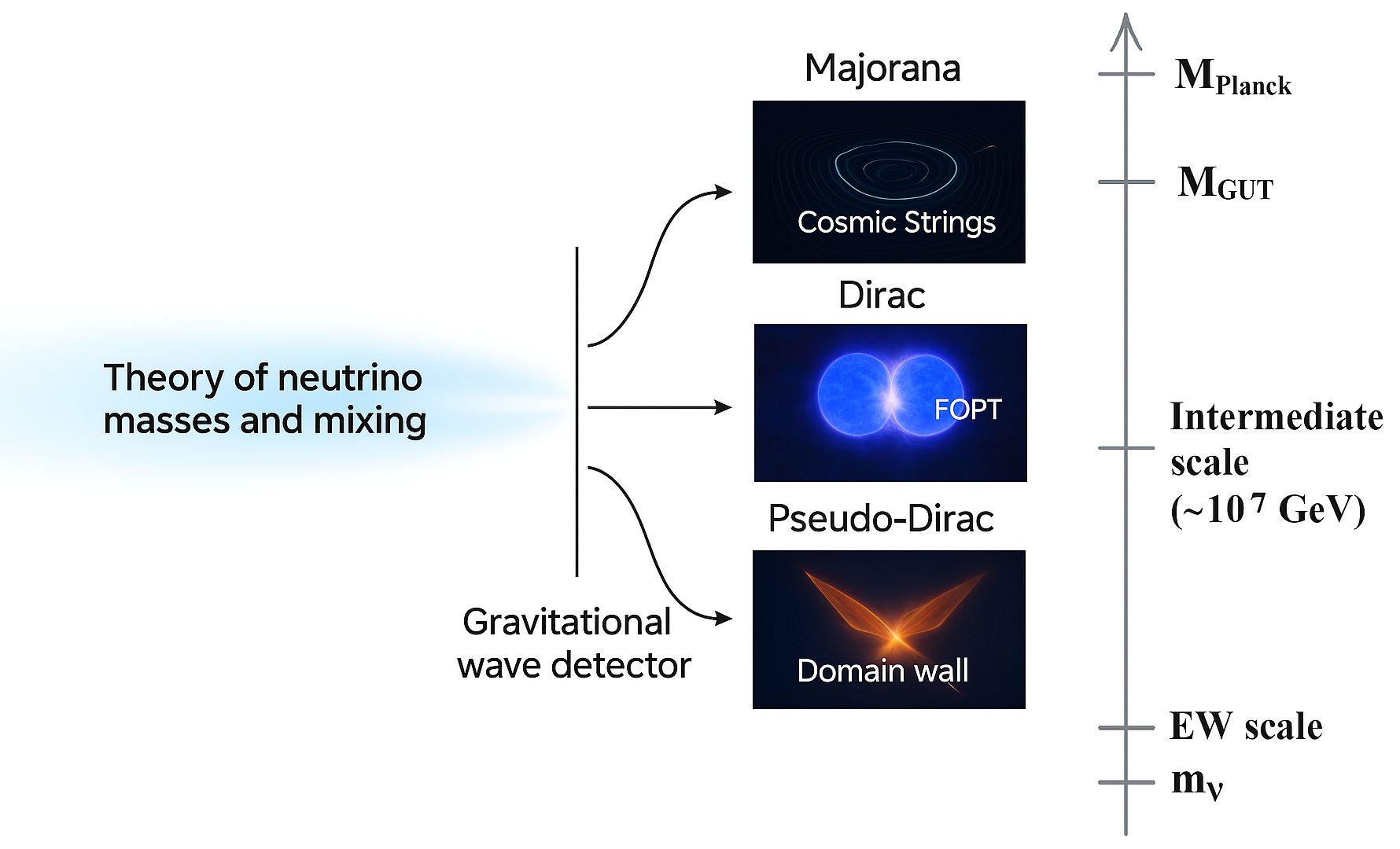}
    \caption{Schematic illustration of characteristic gravitational-wave signatures, depending on whether neutrinos are Majorana, Dirac, or pseudo-Dirac. }
    \label{fig:schem}
\end{figure}

We propose a novel approach rooted in GW phenomenology. Although not all mechanisms of neutrino mass generation produce observable GW signatures, certain scenarios can leave distinctive imprints that reflect the symmetries underlying neutrino masses and mixing. The nature of neutrinos and the origin of their tiny masses are closely connected with status of lepton number global symmetry---conserved for Dirac neutrinos, broken by two units for Majorana neutrinos, and expected to be explicitly violated by quantum-gravity effects unless protected by gauge symmetries. In this context, the most economical option is based on the $B-L$ gauge extension of the SM, which requires only three right-handed neutrinos and a scalar in addition to the SM particle sector.  Importantly, this framework could realize tiny  masses for neutrinos without invoking diminutive values for Yukawa couplings. In this letter, we demonstrate that this framework yields distinctive gravitational-wave spectra for the Dirac, Pseudo-Dirac, and Majorana scenarios: a nearly flat plateau in the Majorana case, a pronounced single peak in the Dirac case, and a characteristic double-peaked structure in the Pseudo-Dirac case (see Fig.~\ref{fig:schem}). Next-generation GW observatories could therefore offer an indirect yet powerful means of resolving the fundamental nature of neutrinos.  

The paper is structured as follows: we first discuss neutrino mass models, underlying symmetries, and their connection to the Dirac, Pseudo-Dirac, and Majorana scenarios within a  $B−L$ extended framework, providing prototypical examples for each. We then analyze the corresponding gravitational-wave signatures arising from these three distinct scenarios. Next, we present our results and discuss the characteristic gravitational-wave imprints associated with the neutrino nature, before concluding.


\begin{figure*}[htb!]
    \centering
    \includegraphics[width=0.85\textwidth]{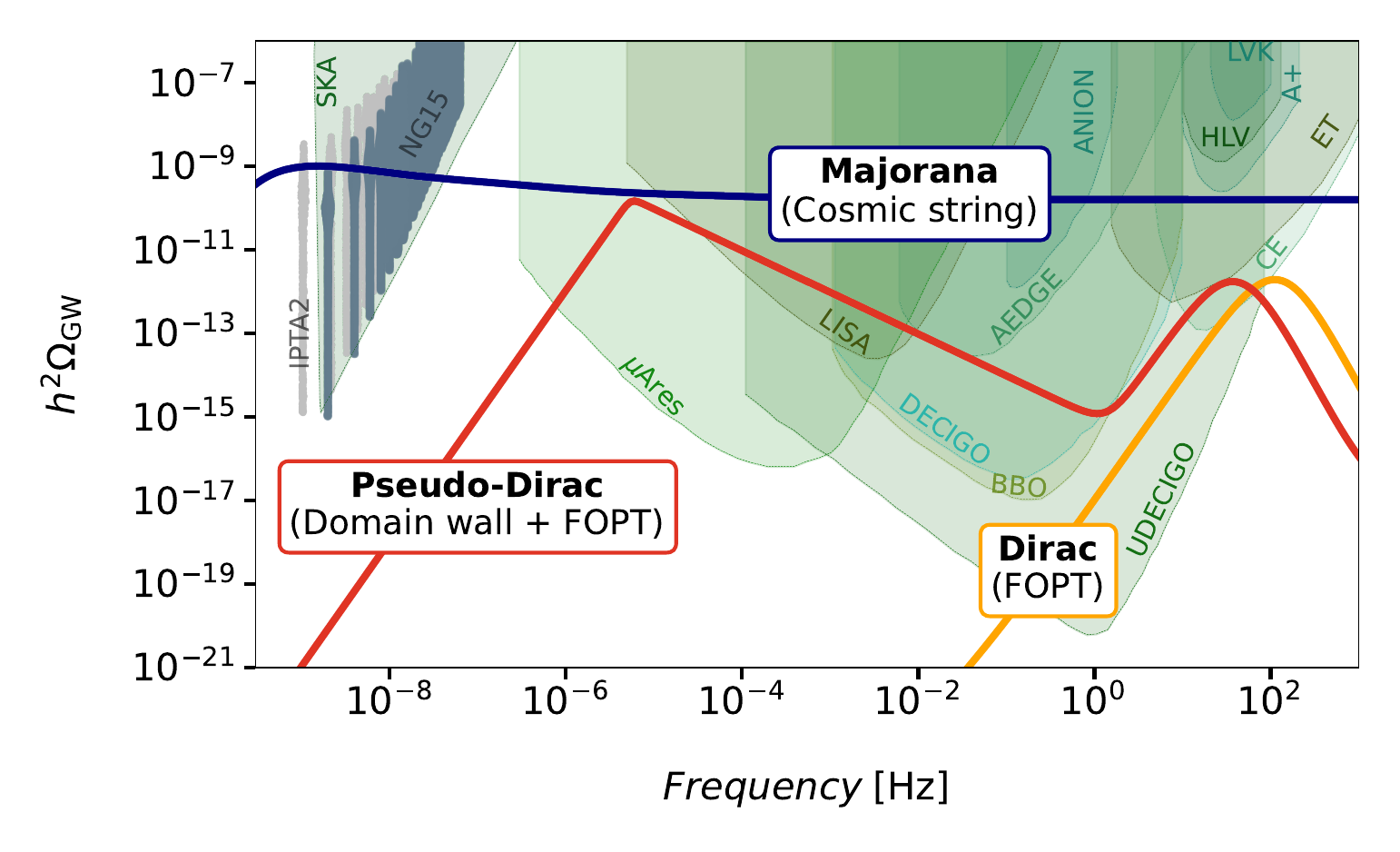}
    \caption{Characteristic gravitational wave spectra for three distinct nature of neutrino masses: Majorana (blue), Dirac (yellow), and pseudo-Dirac (red). The blue curve corresponds to a cosmic string-induced signal with \( v_{BL} = 10^{14} \)~GeV; the yellow represents a FOPT-driven spectrum with \( v_{BL} = 10^{7} \)~GeV; and the red shows the pseudo-Dirac case with \( v_{BL} = 3 \times 10^{6} \)~GeV and \( v_\sigma = 10^{7} \)~GeV. The kink-like feature at low frequencies arises from domain wall annihilation, while the high-frequency peak is sourced by the FOPT. All spectra are computed with \( m_{Z'}/v_{BL} = 0.32 \) and Yukawa coupling fixed at 1.}
    \label{fig:m_plot}
\end{figure*}

\textbf{\emph{Frameworks for Neutrino Mass Generation}.--}
The SM Lagrangian possesses several accidental global symmetries, including lepton number. This symmetry can be broken by a dimension-five operator of the form \( \frac{LLHH}{\Lambda} \), where \( \Lambda \) is the cut-off scale, thereby allowing neutrinos to acquire a Majorana mass. Alternatively, to obtain a purely Dirac nature for neutrinos, one can introduce right-handed neutrinos and generate small Dirac masses via tiny Yukawa couplings; however, this requires the Majorana mass term to be explicitly forbidden. Simply setting it to zero without a protecting symmetry demands justification. A common approach is to impose global lepton number conservation, but such symmetries are generally violated by quantum gravity, which can induce small Majorana masses and render neutrinos pseudo-Dirac \cite{Wolfenstein:1981kw, Petcov:1982ya, Valle:1983dk, Kobayashi:2000md}. A more robust solution is to gauge \( U(1)_{B-L} \), which remains preserved even under quantum gravity, ensuring lepton number conservation and allowing neutrinos to be purely Dirac. In this context, we consider a unified framework that can explain the smallness of neutrino masses for Dirac, pseudo-Dirac, and Majorana cases, without requiring extremely small Yukawa couplings.

The framework is based on $B-L$ gauge extension of the SM, which includes three right-handed neutrinos $N_{R_i}$ (required by gauge-anomaly conditions), a scalar $S$ (for $B-L$ symmetry-breaking), and a gauge boson $Z'$ (associated with $B-L$ symmetry) in addition to the SM particle content. All the BSM states transform (non-)trivially under the ($B-L$) SM gauge symmetry.
With judicious choices of $B-L$ quantum number of the BSM states, the nature of neutrinos and the smallness of their masses can be simultaneously addressed in this setup.

For the scenario of Majorana neutrinos, we consider the following quantum numbers for $N_{R_i}$ and $S$:
\begin{align}
    &\{N_{R_{1,2,3}}\}= (-1,-1,-1);\quad  \{S\}=2.
\label{Eq:ModelMajoranaBL}    
\end{align} 
The relevant Yukawa interactions are given by
\begin{align}
    -\mathcal{L}_{Y}\supset Y_{\nu} \overline{\ell_L} N_R \tilde{H} + \frac{Y_N}{2} \overline{N_R^c} N_R S + h.c.,
\label{eq:ModelMajoranaYuk}   
\end{align}
where $H$ is the SM Higgs field. Both $H$ and $S$ takes non-zero vacuum expectation value (vev): $\langle H \rangle=(0, \frac{v_{EW}}{\sqrt{2}})^T $, $\langle S \rangle= \frac{v_{BL}}{\sqrt{2}}$, where $v_{EW}\simeq 246$ GeV. 
In the limit $Y_N v_{BL}\gg Y_{\nu} v_{EW}$, the mass matrix corresponding to active neutrinos can be approximated as 
\begin{align}
    M_{\nu}\simeq  \frac{v_{EW}^2}{\sqrt{2}v_{BL}}Y_{\nu}Y_N^{-1}Y_{\nu}^T,
\label{eq:ModelMajoranaNeut}    
\end{align}
which corresponds to the standard type-I seesaw mechanism~\cite{Minkowski:1977sc,Yanagida:1979as,Mohapatra:1979ia,Gell-Mann:1979vob,Glashow:1979nm}. 
For simplicity, we ignore neutrino mixing and take $Y_{\nu}\simeq Y_{N}\equiv Y_M$. Then, for neutrino masses of $\mathcal{O}(0.1)$ eV, this yields
\begin{align}
v_{BL}\simeq 10^{14}  \mathrm{GeV} \times Y_M.
\label{eq:ModelMajoranaScale}
\end{align}
As we will discuss later, such large values of $v_{BL}$ lead to a flat GW spectrum generated by the decay of cosmic strings. 

The choice given in Eq.~\eqref{Eq:ModelMajoranaBL} can be modified to suit the scenario of Dirac neutrinos. For instance, consider $\{S\}=3$ instead of $2$ while maintaining the $B-L$ quantum number of $N_{R_i}$ same as in Eq.~\eqref{Eq:ModelMajoranaBL}. In this case, the Majorana mass terms are forbidden to all orders. However, the tree-level Dirac mass terms $Y_{\nu}\overline{\ell_L} N_R \tilde{H}$ are allowed. This implies one requires either unnaturally small Yukawa couplings $Y_{\nu}\simeq \mathcal{O}(10^{-12})$ or fine-tuning in order to comply with neutrino oscillation data. Due to these reasons, we opt for another choice for Dirac neutrinos:
\begin{align}
    &\{N_{R_{1,2,3}}\}= (-4,-4,5);\quad  \{S\}=3.
\end{align} 
Like in the previous case, all the gauge anomaly conditions are satisfied 
for this choice~\cite{Montero:2007cd,Machado:2010ui,Machado:2013oza}. Moreover, the Dirac nature of neutrinos is also protected to all orders~\cite{Ma:2014qra, Ma:2015mjd, Calle:2018ovc, Saad:2019bqf, Jana:2019mgj}. Importantly,  the tree-level Dirac mass terms are also forbidden with this choice and they are generated via Planck-suppressed operators. Consequently, neutrino masses are naturally suppressed in this scenario. The corresponding leading order operator contributing to neutrino mass is given below
\begin{align}
    \mathcal{O}_5\equiv \frac{\overline{\ell_L}\tilde{H}N_{R_{1,2}} S}{M_{\rm Pl}}\Rightarrow M_{\nu}\simeq\frac{v_{EW}v_{BL} Y_D}{2M_{\rm Pl}},
\label{eq:ModelDiracNeut}    
\end{align}
where $M_{\rm Pl}$ is the Planck-scale $\sim 1.22\times 10^{19}$ GeV and $Y_D$ is the coefficients of $\mathcal{O}_5$ operator. Eq.~\eqref{eq:ModelDiracNeut} yields masses for two neutrinos. The third neutrino acquire a tiny mass via a dimension six operator $\overline{\ell_L}\tilde{H}N_{R_{3}} (S^*)^2/M^2_{\rm Pl}$. 
For neutrino masses of $\mathcal{O}(0.1)$ eV, Eq.~\eqref{eq:ModelDiracNeut} implies
\begin{align}
v_{BL}\simeq 10^{7}  \mathrm{GeV} / Y_D.
\label{eq:ModelDiracScale}
\end{align}
As we will show later, in this case the dominant production of GWs is via the first-order phase transition. Consequently, the GW spectrum has a single peak.

We also consider the scenario of pseudo-Dirac neutrinos. To realize this, one requires $M_D\gg M_{L,R}$, where $M_D$ stands for the Dirac neutrino mass and $M_{R (L)}$ denotes the Majorana mass of the right (left)-handed neutrinos. In this limit, the squared mass difference between active and sterile neutrino species can be approximated as $\Delta m^2 \simeq 2M_D M_R$.
The solar neutrino data impose stringent constraints on $\Delta m^2$~\cite{deGouvea:2009fp,Ansarifard:2022kvy}, which translates to $M_D M_R \lesssim 10^{-11}\,\mathrm{eV}^2$.  This implies, for $M_D\simeq 0.1$ eV, the Majorana mass $M_R\lesssim 10^{-10}$ eV. To incorporate these conditions without taking unnaturally small couplings, one must extend the minimal $B-L$ framework. There are multiple ways to do it.  Here, we investigate one such possibility. We focus on a modified version of scenario given in  Eq.~\eqref{Eq:ModelMajoranaBL}, where we consider $\{S\}=\frac{2}{3}$ instead of $2$. In this case, the tree-level Majorana mass terms are forbidden. The dominant contribution to $M_R$ arise via the dimension six operator 
\begin{align}
    &\mathcal{O}_6\equiv \frac{\overline{N_R^c} N_R S^3}{M_{\rm Pl}^2}\Rightarrow M_{R}\simeq\frac{v_{BL}^3 Y_M}{\sqrt{2}M_{\rm Pl}^2},
    \label{eq:Modelpeudo-Dirac}
\end{align}
hence $M_R$ is naturally suppressed in this scenario. Here $Y_M$ is the coefficient of $\mathcal{O}_6$ operator. 
For $M_R\lesssim 10^{-10}$ eV, this implies
\begin{align}
    v_{BL}\lesssim  10^{6} \, \mathrm{GeV}/Y_M^{1/3}.
\end{align}
On the other hand, to forbid the tree-level Dirac mass term, we impose a discrete symmetry $\mathcal{Z}_2$, under which the SM particles and $S$ transform trivially, whereas $N_{R_i}$ transform non-trivially. This symmetry is spontaneously broken by non-zero vev of a scalar field $\sigma$ (denoted by $v_{\sigma}$), which carries non-trivial charge under the $\mathcal{Z}_2$ symmetry only.
Then the leading order contribution to $M_D$ is generated via the  dimension five operator
\begin{align}
        &\mathcal{O}_5\equiv \frac{\overline{\ell_L}\tilde{H}N_R \sigma}{M_{\rm Pl}}\Rightarrow M_{D}\simeq\frac{v_{EW}v_{\sigma} Y_D}{2M_{\rm Pl}}
\end{align}
For $M_D\simeq 0.1$ eV, the above equation implies
\begin{align}
    v_{\sigma}\simeq 10^{7}  \mathrm{GeV}/Y_D.
\end{align}
As we will demonstrate later, this scenario give rise to a double peaked GW spectrum.

\begin{figure}
    \centering
    \includegraphics[width=0.5\textwidth]{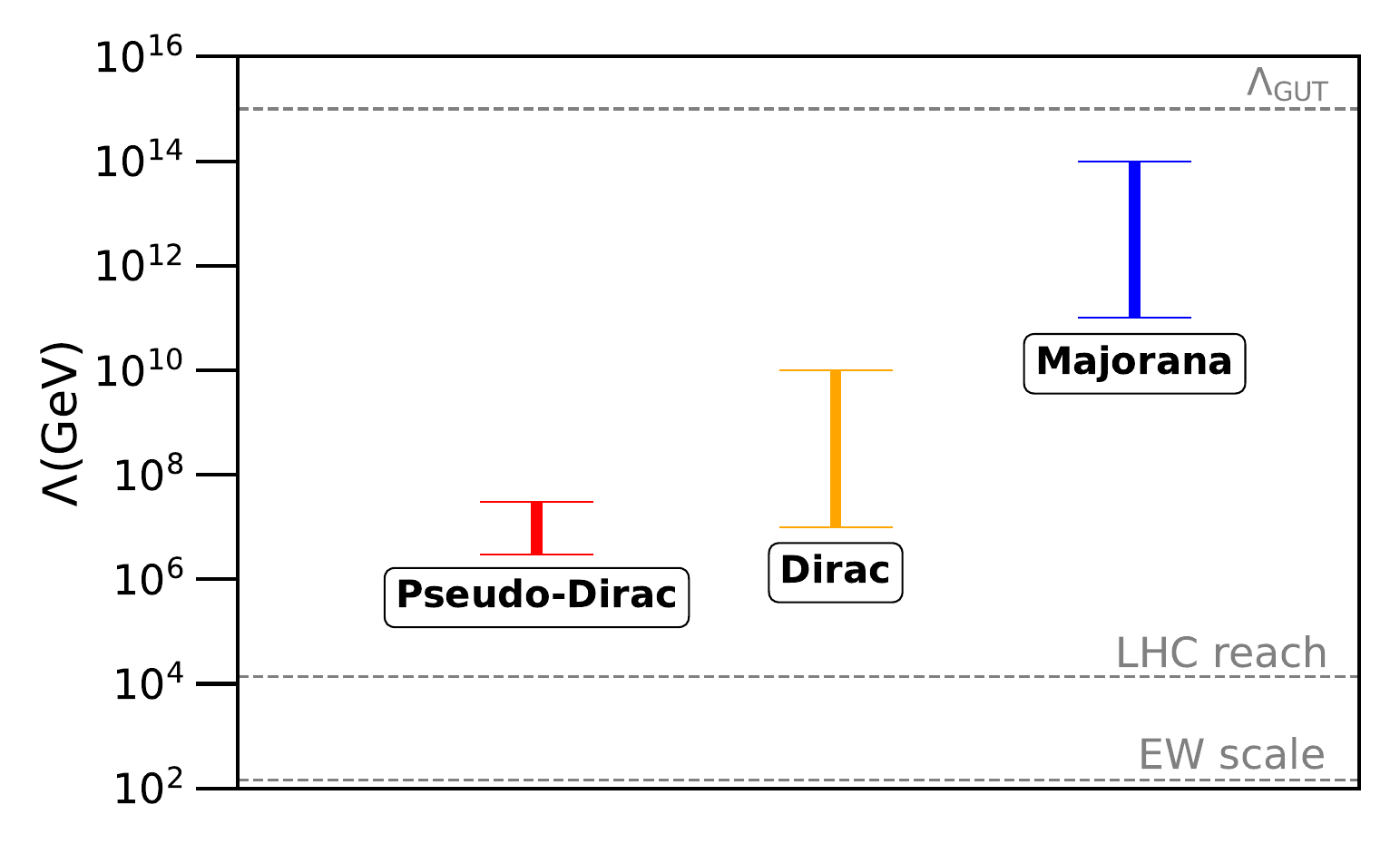}
   \caption{Gravitational wave-sensitive $v_{BL}$ scales for effective Yukawa couplings ($Y_{D/M}$) ranging from $10^{-3}$ to $1$, shown for Dirac, Pseudo-Dirac, and Majorana scenarios (yellow, red, and blue ‘I’s, respectively).}
    \label{fig:BLscale}
\end{figure}

Before discussing the GW spectrum predicted by these scenarios, it is important to highlight their qualitative differences. The main characteristic difference is in the scale of $B-L$ symmetry breaking. In the case of Majorana neutrinos, the symmetry is broken at a high-scale ($\sim 10^{14}$ GeV), whereas for Dirac or pseudo-Dirac neutrinos, the breaking  scale is comparatively low ($\sim 10^{7}$ GeV and $\sim 10^{6}$ GeV, respectively) for Yukawa couplings of $\mathcal{O}(1)$, see Fig.~\ref{fig:BLscale}. Moreover, the realization of pseudo-Dirac neutrinos  also requires spontaneous breaking of  a $\mathcal{Z}_2$ discrete symmetry in our setup. All these features amount to distinctive GW spectrum for the three scenarios that we have considered here, which we discuss next in detail.

\textbf{\emph{GW spectra in the Majorana scenario}.--} We start with the scenario of Majorana neutrinos. In this case, the dominant source of GWs is from the decay of local cosmic strings (LCSs)~\cite{Dror:2019syi,King:2020hyd,Okada:2020vvb,King:2021gmj,Fu:2022lrn,Dasgupta:2022isg,King:2023cgv,Wachter:2024zly,Schmitz:2024gds}. These LCSs are produced as a byproduct of the spontaneous breaking of $B-L$ symmetry, which over the time can form horizon-sized string-network composed of string-segments and string-loops. The total GWs emitted from a loop can be estimated by summing over all its normal modes of frequency $\hat{f}_k=2k/\hat{l}(\hat{t})$, with $\hat{l}(\hat{t})$ being the loop size at the original GW-emission time $\hat{t}$ and $k \in \mathbb{Z}^+$. The corresponding GW-relic density at today can be evaluated by taking into account the redshift and by performing integration over the original emission time, which can be explicitly expressed as
\begin{align}
\Omega_{\text{GW}}^{\text{LCS}}(f) &= 
\sum_k\frac{1}{\rho_c}\frac{2k}{f}\frac{\Gamma_l^{(k)}G\mu^2 F_\alpha}{\alpha\left(\alpha+\Gamma_l G\mu\right)}
\int_{t_F}^{t_0} d\hat{t} \, 
\frac{\tilde{c}\left(t_i^{(k)}\right)}{{t_i^{(k)}}^4} \notag \\
&\quad \times
\theta({t_i^{(k)}}-t_F) 
\left(\frac{a(\hat{t})}{a(t_0)}\right)^5
\left(\frac{a(t_i^{(k)})}{a(\hat{t})}\right)^3,
\label{eq:gw_lcs}
\end{align}
where $\rho_c$ is the critical energy density, $F_\alpha \simeq \alpha =0.1$ denotes a constant, $\tilde{c}$ represents the loop chopping rate~\cite{Blanco-Pillado:2017oxo},  $\mu = \pi v_{BL}^2$  stands for the string-tension (for $v_{BL}=10^{14}$ GeV, $G\mu\sim10^{-10}$) and $\Gamma_l\simeq50$ is a dimensionless quantity related  to GW-emission rate~\cite{Vilenkin:1981bx, Vilenkin:2000jqa}. The parameter $t_F$ indicates loop formation time and the quantity $t_i^{(k)}$ is parametrized as 
\begin{align}
 t_i^{(k)}(\hat{t},f)=\frac{(\hat{l}(\hat{t},f,k)+\Gamma_l G \mu \hat{t})}{( \alpha+\Gamma_l G \mu)}.
    \label{Eq:t_hat}
\end{align}
Here, $\Gamma_l^{(k)} \sim k^{-r}$ and $r = [2, 4/3]$ refer to GW radiation from kink-kink collisions and from cusps (local defects in the loops), respectively. Evaluating $\Omega_{\text{GW}}^{\text{LCS}}(f)$ is computationally expensive task for larger $k$. Therefore, in this study we adopt a simulated template named \texttt{stable-c} (available in \texttt{PTArcade} software~\cite{Mitridate:2023oar}) and assumes $r=4/3$ ~\cite{Vachaspati:1984gt}.

\begin{figure}
    \centering
    \includegraphics[width=0.5\textwidth]{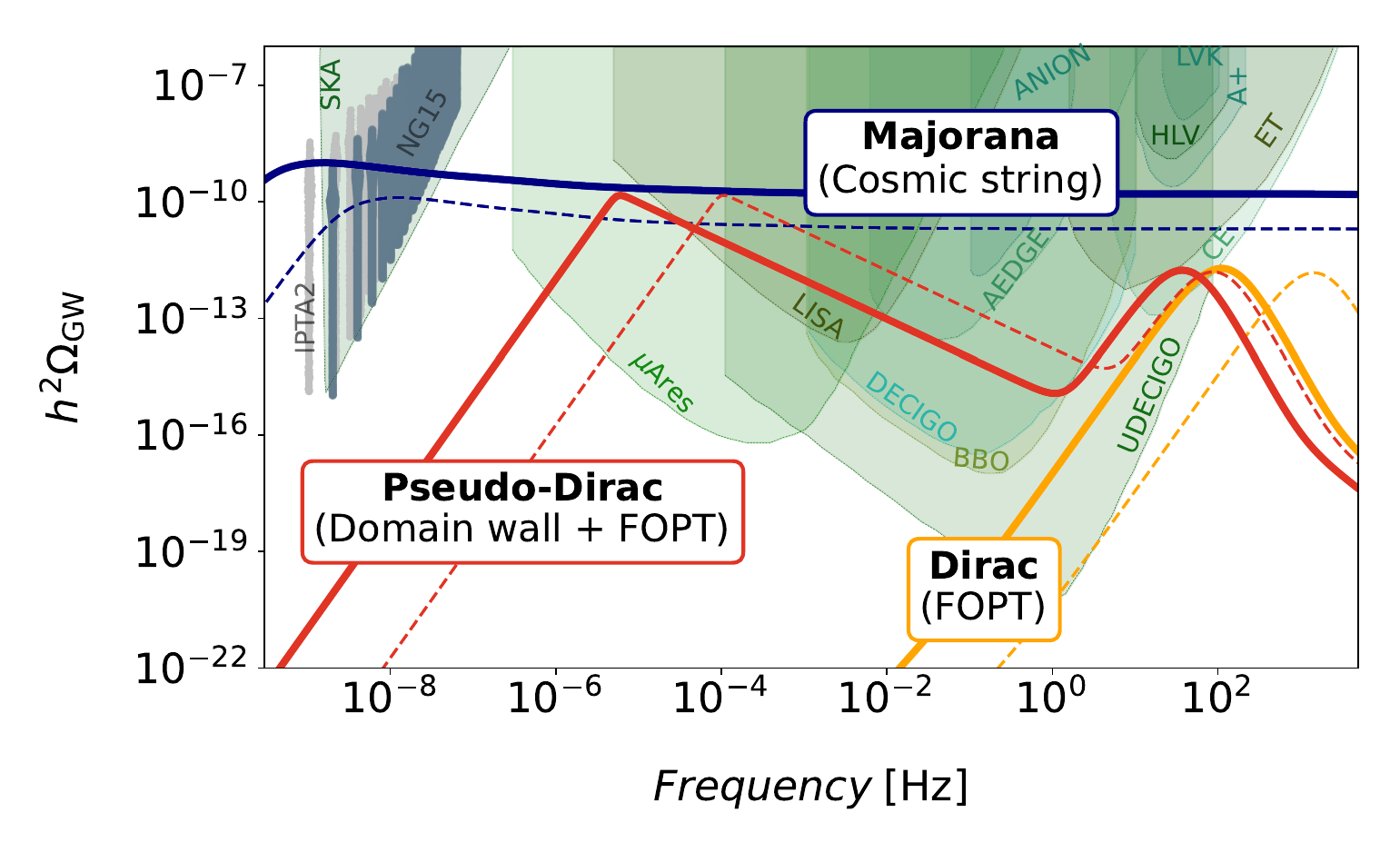}
    \caption{ The blue, yellow and red solid (dashed) lines represent the gravitational wave spectrum correspond to Majorana, Dirac and pseudo-Dirac scenarios, respectively for Yukawa coupling $Y_{D/M}\sim 1$ ($ 0.1$).}
    \label{fig:diffYukawas}
\end{figure}

The resulting GW spectrum for $Y_M=1$ ($v_{BL}\simeq 10^{14}$ GeV) is shown in  Fig.~\ref{fig:m_plot} as a blue solid line (also see Fig.~\ref{fig:diffYukawas}). Notice that the spectrum has a nearly flat shape across a wide range of frequencies. Moreover, the amplitude of the spectrum  lies within the projected sensitivity of future ground and space based detectors- $\mu$Ares~\cite{Sesana:2019vho}, LISA~\cite{LISA:2017pwj}, DECIGO~\cite{Kawamura:2020pcg}, UDECIGO~\cite{Kudoh:2005as}, BBO~\cite{Harry:2006fi}, AEDGE~\cite{AEDGE:2019nxb}, Einstein Telescope (ET)~\cite{Hild:2008ng}, Cosmic Explorer (CE)~\cite{LIGOScientific:2016wof},  ANION~\cite{Badurina:2019hst}, and SKA~\cite{Janssen:2014dka}. The results from the IPTA2 (IPTA second data release)~\cite{Antoniadis:2022pcn} and NG15 (NANOGrav 15 year data)~\cite{NANOGrav:2023gor,NANOGrav:2023hde,NANOGrav:2023hvm} are also shown in Fig.~\ref{fig:m_plot} \footnote{Particularly in~\cite{NANOGrav:2023hvm}, the working group mentioned about this LCSs as a potential source of this NG15 violins. However, to justify CSs as the true source of this violins we have to wait for more data.}.

In addition to LCSs, the first-order phase transition (FOPT) could also induce GWs in this setup~\cite{Hasegawa:2019amx}. However, we find that the peak frequency of the corresponding spectrum lies well above kilo-hertz range for  $Y_M\gtrsim 10^{-6}$, 
hence falls beyond the reach of
next generation GW experiments. Moreover, the amplitude of the spectrum is  found to be comparatively smaller than that of LCSs driven spectrum for  scenarios where Yukawa couplings are not small. Consequently, in this case, the GW spectrum is fully driven by LCSs.

\textbf{\emph{GW spectra in the Dirac scenario}.--} Next, we consider the scenario of Dirac neutrinos. As aforementioned, in this case the $B-L$ symmetry is broken comparatively at a  low scale $\sim 10^7$ GeV, which leads to a very different phenomenology for GWs compared to the Majorana scenario.
Below we discuss these phenomenological differences. 

Like in the case of Majorana neutrinos, both LCSs and FOPT can induce GWs in this scenario. However, we find that the GW spectrum driven by LCSs falls well below the sensitivity of future GW detectors. On the other hand, the FOPT can yields a sufficient contribution to $\Omega_{\rm GW}$ and hence it can serve as an indicator for the scenario of Dirac neutrinos. 

To estimate the corresponding GW spectra,  first we focus on the finite temperature corrections to the scalar potential, which dictates the phase transition. Since $v_{BL}\gg v_{EW}$, it is sufficient to study the scalar potential of $S$ only~\cite{Hasegawa:2019amx,Costa:2025csj}. The  corresponding one-loop effective potential  at a finite temperature $T$ is given by~\cite{Carrington:1991hz} (using Arnold-Espinosa method)
\begin{align}
V(S,T) = V_{tree}(S) + V_{CW}(S) + V_{th}(S,T) + V_{daisy}(S,T),
\label{Eq:V_phi}
\end{align}
where $V_{tree}$  stands for the tree-level potential,  $V_{CW}(S)$ denotes the one-loop ($T=0$) Coleman-Weinberg potential, while $V_{th}(S,T) + V_{daisy}(S,T)$ represent $T\neq0$ corrections. 
The  explicit form these terms are given in the End Matter. For $B-L$ gauge coupling $g_{BL}\gtrsim \mathcal{O}(0.1)$, this scenario yields a FOPT~\cite{Hasegawa:2019amx}.

During a FOPT, there are mainly three sources that contributes to GWs: collision of nucleated bubbles, sound waves exerted due to bubble collision~\cite{Hindmarsh:2013xza} and turbulence~\cite{Kamionkowski:1993fg}. Their contributions can be parametrized in terms of $\alpha$, $\beta/H_*$ and $T_*$. Here, $\alpha$ determine the strength of the FOPT,  $\beta/H_*$ measures the inverse time scale of the FOPT, while the quantity $T_*$ denotes the nucleation temperature. The explicit forms of the parameters are given in the End Matter. For estimating these parameters, we employ the scalar potential in  \texttt{CosmoTransitions} software~\cite{Wainwright:2011kj}. We obtain $\alpha \sim 1$, $\beta/H_*\in \mathcal{O}(10^3-10^4)$, and $T_*\in \mathcal{O}(10^5-10^6)$ GeV for the benchmark values given in Tab.~\ref{benchmarkpoints_FOPT_lam_1}. Since $\alpha \sim 1$, the contributions to GWs from the bubble collision is negligible in our scenario~\cite{Ellis:2019oqb}. 
On the other hand, both sound waves and turbulence contributes significantly to GW signals, however, the later primarily affects  the high-frequency tail of spectra only.
Hence, here, we mainly focus on the contributions from sound waves.

The GW-relic density from sound waves can be evaluated by  ~\cite{Caprini:2015zlo} 
\begin{align}
\Omega^{\text{sw}}_{\text{GW}}(f)\simeq\Omega^{\text{sw}}_{\text{peak}}\left(\frac{f}{f_{\text{peak}}^{\text{sw}}}\right)^3\left(\frac{7}{4+3(\frac{f}{f_{\text{peak}}^{\text{sw}}})^2})\right)^{7/2}.
\label{Eq:Omega_sw}
\end{align}
Here, $\Omega^{\text{sw}}_{\text{peak}}$ and $f_{\text{peak}}^{\text{sw}}$ represent the peak amplitude and peak frequency of the sound wave driven GW spectrum, respectively. These quantities can be determined by ~\cite{Caprini:2015zlo} (assuming unit bubble wall velocity)
\begin{align}
h^2\Omega^{\text{sw}}_{\text{peak}}\simeq2.65\times 10^{-6}\kappa_v^2\Upsilon\left(\frac{H_*}{\beta}\right)\left(\frac{\alpha}{1+\alpha}\right)^2\left(\frac{g_*}{100}\right)^{-\frac{1}{3}},
\label{Eq:Omega_sw_peak}
\end{align}
\begin{align}
f_{\text{peak}}^{\text{sw}}\simeq19\left(\frac{\beta}{H_*}\right)\left(\frac{T_*}{10^8 \text{GeV}}\right)\left(\frac{g_*}{100}\right)^{1/6} \text{Hz},
\label{Eq:f_sw_peak}
\end{align} 
where $g_*$ is the number of relativistic degrees of freedom and the parameter, $\kappa_v$ has the form, $\kappa_v=\alpha/(0.73+0.083\sqrt{\alpha}+\alpha)$~\cite{Espinosa:2010hh}. The term, $\Upsilon$ denotes the suppression factor that arises due to the finite lifetime of sound waves~\cite{Guo:2020grp}. We find its value $\mathcal{O}(0.01)$ for the benchmark values given in Tab.~\ref{benchmarkpoints_FOPT_lam_1}.  

\begin{table}[th]
\centering
\begin{ruledtabular}
\renewcommand{\arraystretch}{1.3}  
\begin{tabular}{ccccccc}
$Y_D$ & $ v_{BL}$ (GeV) & $\alpha$ &
 $\beta/H_*$  & $c_s^2$ & $T_c$ (GeV) &
$T_*$ (GeV)   \\
\hline 
 1.0 & $10^{7}$  & 1.5233 & 2160 & 0.3331 &  1042502 & 274622 \\ \hline
 0.1 & $10^{8}$  & 1.5233 & 2697 & 0.3331 &  10425066 & 2746331 \\ \hline
 1.0 & $3\times 10^6$  &  1.5235 & 2342 & 0.3331  & 312750 & 82384 \\ \hline
 0.1 & $6.46\times 10^6$  &   1.5234 &  2584 & 0.3331  & 667201 & 175756 
\\
\end{tabular}
\end{ruledtabular}
\caption{
\label{benchmarkpoints_FOPT_lam_1} Benchmark values of the parameters for FOPT in the Dirac scenario (first two rows) and in the pseudo-Dirac scenario (last two rows) for $m_{Z'}/v_{BL}=0.32$ and $\lambda_S=2.16 \times 10^{-4}$. $c_s$ and $T_c$ denote the sound-wave velocity and the critical temperature of the FOPT, respectively, obtained using \texttt{CosmoTransitions}.
}
\end{table}


The estimated GW spectrum  for Yukawa coupling $Y_D=1$ ($v_{BL}=10^7$ GeV) is  shown in Fig.~\ref{fig:m_plot} as a yellow solid line. 
Here, we fix the gauge coupling through $m_{Z'}/v_{BL}=0.32$ ($m_{Z'}$ denotes the mass of the $B-L$ gauge boson) and also set the quartic coupling $\lambda_S=2.16 \times 10^{-4}$ (see the End Matter). The corresponding benchmark values of the FOPT parameters are given in Tab.~\ref{benchmarkpoints_FOPT_lam_1}. 
Notice that the spectrum features a pronounced single peak with $h^2\Omega^{\text{sw}}_{\text{peak}}\simeq10^{-12}$ and $f_{\text{peak}}^{\text{sw}}\sim$ hecto-hertz, consequently, it is well within the reach of next generation GW detectors, such as UDECIGO and CE. The corresponding spectra for smaller values of Yukawa couplings are given in Fig.~\ref{fig:diffYukawas}. As it can be seen, the spectrum shift towards the high-frequency regime for smaller Yukawa couplings.

\textbf{\emph{GW spectra in the Pseudo-Dirac scenario}.--}
We now turn our attention to the case of pseudo-Dirac neutrinos. The realization of this scenario requires spontaneous breaking of both $B-L$ gauge symmetry and a discrete symmetry $\mathcal{Z}_2$. The impact of the former on the GW spectra is similar to that of in Dirac scenario due to a comparable value of $v_{BL}$ in both setups. On the other hand,  the spontaneous breaking of the discrete symmetry is a distinct feature of pseudo-Dirac neutrinos, which leads to characteristic differences in GW phenomenology compare to Majorana and Dirac cases. We examine these differences in detail below.

For the analysis, we assumes the tree-level scalar potential to be $V(\sigma)=\lambda_\sigma (\sigma^2- v_\sigma^2)^2/4$ \footnote{If we introduce $\lambda_{s\sigma}(S^*S)\sigma^2$ term in the Lagrangian, similar DW-phenomenology will be there as like the $\lambda_{s\sigma}=0$ scenario. However, the FOPT could be affected depending on the value of $\lambda_{s\sigma}$ (and may occur for different benchmark values of the model parameters compare to the $\lambda_{s\sigma}=0$ benchmarks). In addition, since $v_{EW}\ll v_S,v_\sigma$, we are ignoring all interaction terms between SM and new physics scalars $S$, $\sigma$ for this scenario.}. The non-zero vev of $\sigma$ breaks the $\mathcal{Z}_2$ symmetry spontaneously, leading to the formation of domain walls (DWs). If such DWs are long lived, they would eventually dictate the Universe's total energy density~\cite{Zeldovich:1974uw}. Such scenarios are incompatible with standard cosmology~\cite{Planck:2013pxb}. To avoid this, we introduce a soft $\mathcal{Z}_2$ breaking term $\delta{V}(\sigma)=\epsilon\sigma v_\sigma\left({\sigma^2}/{3}-v_\sigma^2\right)$ to $V(\sigma)$, which in effect generates an asymmetry between the heights of the potential at $\sigma=\pm v_{\sigma}$, $V_{bias} =(4/3)\epsilon v_\sigma^4$. 
As a consequence of this, the DWs become unstable, leading to their decay and annihilation. 
 
Presuming this annihilation happens instantaneously, the  contribution to the GW amplitude can be determined by~\cite{Caprini:2019egz}
\begin{align}
h^2\Omega^{\text{DW}}_{\text{GW}}(f)\simeq\Omega^{\text{peak}}_{\text{DW}}h^2  \frac{(3 + b)^c}{\left[ b \left( \frac{f}{f_{\text{peak}}} \right)^{-3/c} + 3 \left( \frac{f}{f_{\text{peak}}} \right)^{b/c} \right]^c},
\end{align}
where the parameters b and c are allowed to be in the range $\left[0.5, 1\right]$ and $\left[0.3, 3\right]$, respectively~\cite{Hiramatsu:2010yz,Ferreira:2022zzo}. The quantities $\Omega^{\text{DW}}_{\text{peak}}$ and $f_{\text{peak}}^{\text{DW}}$ can be estimated by~\cite{Saikawa:2017hiv}
\begin{align}
h^2\Omega^{\text{DW}}_{\text{peak}} &\simeq 1.49 \times 10^{-10}  \left( \frac{10.75}{g_*} \right)^\frac{1}{3}  \left( \frac{10^7\, \text{GeV}^4}{V_{\text{bias}}} \right)^2 \notag \\
&\quad \times 
\left( \frac{\mathcal{E}^{1/3}}{10^7\, \text{GeV}} \right)^{12}
\label{Eq:DW_GW_peak}
\end{align}
and
\begin{align}
f_{\text{peak}}^{\text{DW}} \simeq 5.93 \times 10^{-6} \left( \frac{\mathcal{E}^{1/3}}{10^7\, \text{GeV}} \right)^{-\frac{3}{2}} \left( \frac{V_{\text{bias}}}{10^7~\text{GeV}^4} \right)^\frac{1}{2}.
\label{Eq:DW_GW}
\end{align}
Here $\mathcal{E} \simeq \sqrt{8\lambda_\sigma}v_\sigma^3/3$ denotes the DW's surface tension.

In Fig.~\ref{fig:m_plot}, we show the predicted GW spectrum for Yukawa couplings $Y_{D/M}=1$ as a red solid line. The double-peaked nature of the spectrum is a characteristic feature of pseudo-Dirac scenario. Here, the peak in the low (high) frequency regime arises from DW annihilation (FOPT)\footnote{We disagree with the conclusion of Ref.~\cite{King:2023cgv}, which associates a kink-like gravitational wave spectrum with the Dirac neutrino scenario. We show that such a feature more naturally arises in the pseudo-Dirac case, where tiny Majorana masses are generated via Planck-suppressed operators from quantum gravity. In contrast, a genuinely Dirac scenario in our framework yields a sharp peak-like spectrum.}. Remarkably, the amplitude and frequency of the peaks lie within the reach of various next generation GW experiments, such as $\mu$Ares, LISA, DECIGO, UDECIGO, BBO, AEDGE and CE. See Fig.~\ref{fig:diffYukawas} for corresponding spectra for smaller values of Yukawa couplings. For generating these plots, we use the benchmark values given in Tabs.~\ref{benchmarkpoints_FOPT_lam_1} and \ref{benchmarkpoints_DW}.

\begin{table}[th]
\centering
\begin{ruledtabular}
\renewcommand{\arraystretch}{1.3}  
\begin{tabular}{cccc}
$Y_M$ & $ v_\sigma$ & $\lambda_\sigma$ &$V_{bias}$ \\
\hline 
 1.0 & $10^{7}$ GeV &1& $10^7 $ $\text{GeV}^4$ \\ \hline
 0.1 & $10^{8}$ GeV &1& $10^{13} $ $\text{GeV}^4$ 
\\
\end{tabular}
\end{ruledtabular}
\caption{
\label{benchmarkpoints_DW} Benchmark values of the parameters for the $Z_2$ DW annihilation in the pseudo-Dirac scenario.
}
\end{table}
\vspace{0.1in}

\textbf{\emph{Conclusion}.--}
In this letter, we have demonstrated that the next generation gravitational wave searches could shed light on the nature of neutrino masses. 
To illustrate this, we have considered  an economical framework based on the $B-L$ gauge extension of the SM that yield naturally light neutrino masses without invoking tiny values for Yukawa couplings.
We show that this setup predicts distinctive gravitational wave spectra for Majorana (a nearly flat plateau driven by LCSs), Dirac (a pronounced single peak generated by FOPT), and Pseudo-Dirac scenarios (a characteristic double-peaked structure induced by DW+FOPT)
in a way that they could be distinguished in next generation gravitational wave observatories. 
Thus, a characteristic gravitational wave signature could offer a unique window into the origin  of neutrino masses and unveil the fundamental symmetry governing their nature.

\vspace{0.1in}
\begin{acknowledgments}
{\textbf {\textit {Acknowledgments.--}}} SM would like to thank Utsav Atta for useful discussions.
SJ and SM would like to acknowledge the support from the Department of Atomic Energy (DAE), Government of India, for the Regional Centre for Accelerator-based Particle Physics (RECAPP), Harish-Chandra Research Institute.
\end{acknowledgments}

\section*{End Matter}
\textbf{\emph{Form of $V(S)$ at finite temperature}.--}
To estimate the components of $V(S, T)$ (which we introduce in Eq.~\ref{Eq:V_phi}), we first replace $S$ by classical background field $\mathcal{S}/\sqrt{2}$ and use the underlying equations~\cite{Arnold:1992rz,Quiros:1999jp},
\begin{align}
V_{tree}(\mathcal{S})=-\frac{1}{2}\mu_S^2 \mathcal{S}^2+\frac{1}{4}\lambda_S \mathcal{S}^4,
\label{Eq:V_tree}
\end{align}

\begin{align}
V_{\text{CW}}(\mathcal{S}) = \frac{1}{64\pi^2} 
&\sum_i n_i\Bigg[  2 m_i^2(\mathcal{S}) m_i^2(v_{BL}) \notag \\
&\ + m_i^4(\mathcal{S}) \left( \log \frac{m_i^2(\mathcal{S})}{m_i^2(v_{BL})} - \frac{3}{2} \right)
\Bigg],
\label{Eq:V_CW}
\end{align}
Here, $V_{tree}(\mathcal{S})$ defines the tree level potential and $V_{\text{CW}}(\mathcal{S})$ is the one-loop zero temperature CW-contribution to $V(\mathcal{S}, T)$. The index, $i$ includes the scalar $S$, the Goldstone boson $\chi$, the gauge boson $Z'$ and 3 Dirac neutrinos with field dependent masses $m_{S}^2(\mathcal{S})=3\lambda_S \mathcal{S}^2-\mu_S^2$, $m_{\chi}^2(\mathcal{S})=\lambda_S \mathcal{S}^2-\mu_S^2$ and $m_Z'(\mathcal{S})=|s|g_{BL} \mathcal{S}$, respectively, where $s$ is the $U(1)_{B-L}$ charge of $S$. The mass terms for neutrino species are given in Eq.~(\ref{eq:ModelDiracNeut}) and Eq.~(\ref{eq:Modelpeudo-Dirac}) for Dirac and pseudo-Dirac scenarios respectively. The parameter $n_i$ denotes the number of degrees of freedom of species $i$ ($n_{Z'}=3, n_S=1, n_\chi=1$ and ${n_\nu}_i=-4$ and $-2$ for Dirac and Pseudo-Dirac scenarios, respectively).

The other two contributions ($V_{th}(\mathcal{S},T)$ and $V_{daisy}(\mathcal{S}, T)$) are temperature dependent and we express them as follows~\cite{Carrington:1991hz},
\begin{align}
    V_{th}(\mathcal{S},T)= \frac{T^4}{2\pi^2}\Bigg[
    &\ \sum_{i = \text{bosons}} n_i J_B\left( \frac{m_i^2(\mathcal{S})}{T^2} \right) \notag \\
&\ 
+ \sum_{i = \text{fermions}} n_i J_F\left( \frac{m_i^2(\mathcal{S})}{T^2} \right) \bigg].
    \label{Eq:V_th}
\end{align}
 Here the function  $J_{F(B)}$ for the fermions (bosons) is represented by
\begin{align}
    J_{F(B)}(x) = \ \int_0^{\infty} dy\, \Bigg[ y^2\log\left(1 \pm e^{-\sqrt{x^2 + y^2}} \right)\Bigg],
    \label{Eq:jb/f}
\end{align}
and,
\begin{align}
    V_{daisy}(\mathcal{S}, T) = \frac{T}{12\pi} &\ \sum_{i = \text{bosons}} \tilde{n}_i \Bigg[
     \left(m_i^2(\mathcal{S})\right)^{3/2} \notag \\
&\ 
 - \left(m_i^2(\mathcal{S}) + \Pi_i(T) \right)^{3/2} \Bigg],
\label{Eq:V_daisy}
\end{align}
where $\tilde{n}_i=1$ for $S, \chi$ and $Z'$, since out of 3 degrees of freedom of $Z'$ ($n_{Z'}=3$) only the longitudinal component receive the thermal mass. Here, $\Pi_i(T)$ stands for the high temperature self energy corrections to these species and their forms are given below,
\begin{equation}
\begin{aligned}
    \Pi_S(T) = \left( \frac{1}{4} \left(\frac{m_{Z'}}{v_{BL}}\right)^2 + \frac{\lambda_S}{3} \right) T^2 , \\
    \Pi_{Z'}(T)\big|_{\text{Dirac}} = \frac{13}{9}\, \left(\frac{m_{Z'}}{v_{BL}}\right)^2 T^2 , \\
    \Pi_{Z'}(T)\big|_{\text{Pseudo-Dirac}} = \frac{35}{24} \left(\frac{m_{Z'}}{v_{BL}}\right)^2 T^2 .
\end{aligned}
\label{Eq:Pi_i}
\end{equation}
Here, we neglect the effective Dirac Yukawa coupling between neutrinos and $S$ in the $\Pi_S(T)$ expression (see Eq.~(\ref{eq:ModelDiracNeut}) and Eq.~(\ref{eq:Modelpeudo-Dirac}) for details), since it is $M_{Pl}$ suppressed.

\vspace{1cm}
\textbf{\emph{Form of FOPT parameters $\alpha$, $\beta/H_*$}.--}  
The quantity $\alpha$ is  defined using the pseudotrace as~\cite{Giese:2020rtr,Athron:2023xlk}
\begin{align}
\alpha=\frac{\Delta\left[V-T\frac{\partial V}{\partial T}+\frac{V}{c_s^2}\right]_{T_*} }{\big[-3T\frac{\partial V}{\partial T}\big]_{S_+, T_*}},
\label{Eq:alpha}
\end{align}
where, $\Delta[f]_{T_*} \equiv f(S_+,T_*) - f(S_-,T_*)$ and $S_{+(-)}$ corresponds to $S$-value at false (true) minima and the term $c_s^2(=[(1/T)(\partial V/\partial T)/(\partial^2 V/\partial T^2)]_{S-,T_*}$), stands for the square of sound wave velocity at true minima, while the parameter $\beta/H_*$ is evaluated using  
\begin{align}
\frac{\beta}{H_*}\simeq T\frac{d(S^3_E/T)}{dT}\Bigg|_{T_*},
\label{Eq:beta}
\end{align}
where $S^3_E$ denotes the 3-dimensional Euclidean action~\cite{Turner:1992tz}.

\vspace{1cm}
\textbf{\emph{ The turbulence-driven Gravitational wave spectra}.--}  
In general, the amplitude of the full GW-spectrum is mostly controlled by  $\Omega^{\text{sw}}_{\text{GW}}(f)$. However, the high frequency regime of the spectrum usually have the turbulence dependence. 
In a similar fashion like sound wave, the peak amplitude and the frequency of the spectrum originated by the turbulence are approximated as~\cite{Kamionkowski:1993fg}
\begin{align}
h^2\Omega^{\text{tur}}_{\text{peak}}\simeq3.4\times 10^{-4}\left(\frac{H_*}{\beta}\right)\left(\frac{\kappa_{tur}\alpha}{1+\alpha}\right)^\frac{3}{2}\left(\frac{g_*}{100}\right)^{-\frac{1}{3}},
\label{Eq:Omega_tur_peak}
\end{align}
\begin{align}
f_{\text{peak}}^{\text{tur}}\simeq27\left(\frac{\beta}{H_*}\right)\left(\frac{T_*}{10^8 \text{GeV}}\right)\left(\frac{g_*}{100}\right)^{1/6} \text{Hz},
\label{Eq:f_tur_peak}
\end{align}
where $\kappa_{tur}=0.05\kappa_v$. The turbulence's contribution to the spectrum in terms of $\Omega^{\text{tur}}_{\text{peak}}$ and $f_{\text{peak}}^{\text{tur}}$ takes the form~\cite{Caprini:2009yp,Caprini:2015zlo,Jinno:2015doa}
\begin{align}
\Omega^{\text{tur}}_{\text{GW}}(f)\simeq\Omega^{\text{tur}}_{\text{peak}}\left(\frac{f}{f_{\text{peak}}^{\text{tur}}}\right)^3 \frac{\left(1+ \frac{8\pi f}{h_*}\right)^{-1}}{(1+\frac{f}{f_{\text{peak}}^{\text{tur}}})^{11/3}}.
\label{Eq:Omega_tur}
\end{align}
with 
\begin{align}
h_*=17\left(\frac{g_*}{100}\right)^\frac{1}{6}\left(\frac{T_*}{10^8 \text{GeV}}\right) \text{Hz}.
\label{Eq:h_star}
\end{align}

\begin{figure}
    \centering
    \includegraphics[width=0.5\textwidth]{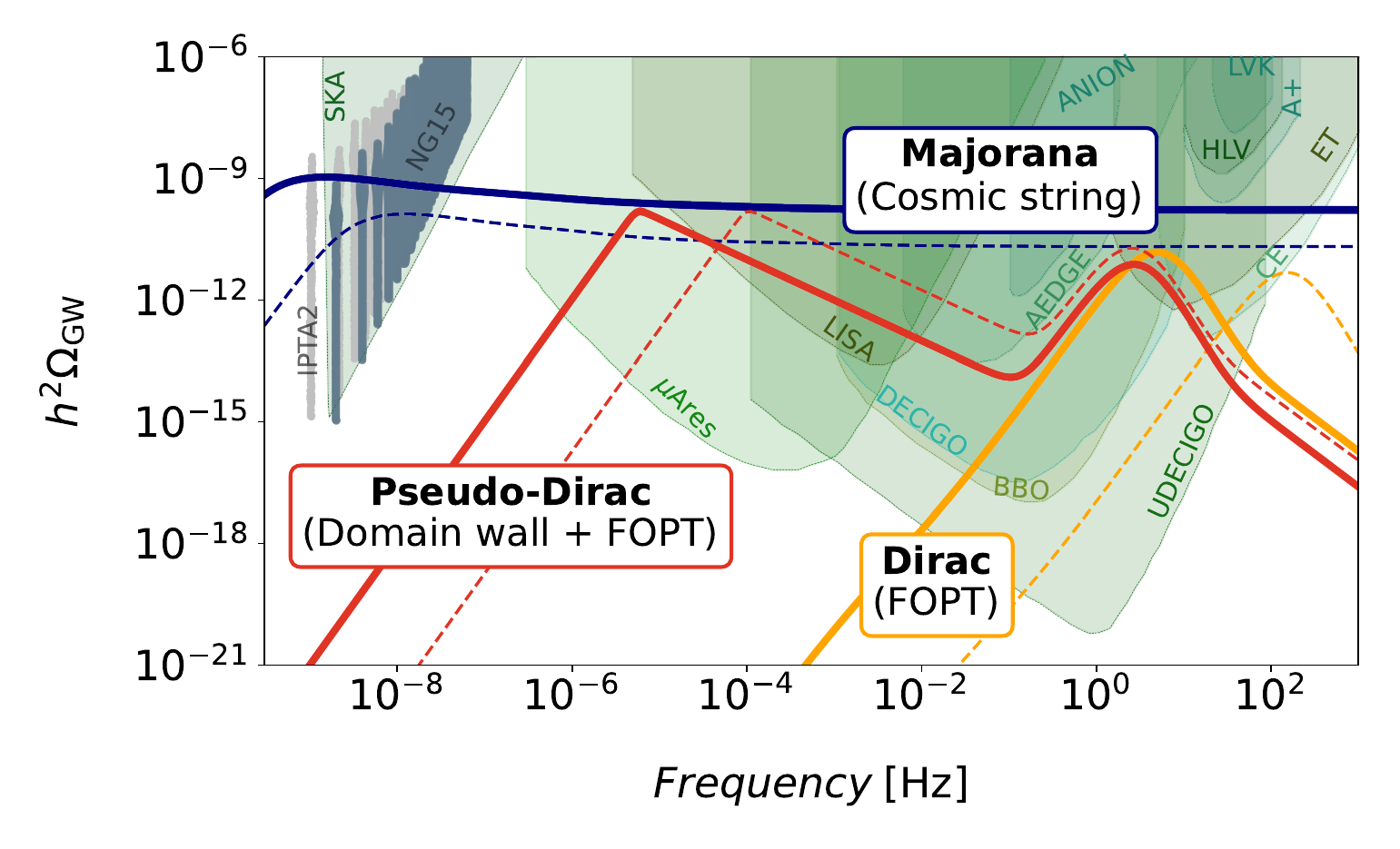}
    \caption{The plot shows the gravitational wave spectra for Dirac, Majorana and Pseudo-Dirac scenarios for the benchmarks given in  Tab.~\ref{benchmarkpoints_DW} and Tab.~\ref{benchmarkpoints_FOPT_lam_1_new}. Here, all the colors of the solid (dashed) lines represent the same as like in Fig.~\ref{fig:diffYukawas}, where the Majorana spectrum (blue, solid and dashed) is shown without modifications. }
    \label{fig:m_plot_new}
\end{figure}

\vspace{1cm}
\textbf{\emph{Different set of benchmark values}.--} For benchmark values given in Tab.~\ref{benchmarkpoints_FOPT_lam_1}, we fix the quartic coupling $\lambda_S=2.16 \times 10^{-4}$. Here, we consider varying $\lambda_S$,
which allows for the possibility of relatively smaller values of $T_*$ and $\beta/H_*$ compared to those in Tab.~\ref{benchmarkpoints_FOPT_lam_1} (one may find this for another fixed value of $\lambda_S$ as well); hence, yielding a better detection probability for GW spectra. One such scenario is given in Tab.~\ref{benchmarkpoints_FOPT_lam_1_new}  and the corresponding GW spectra is displayed in Fig.~\ref{fig:m_plot_new}.

\begin{table}[th]
\centering
\begin{ruledtabular}
\renewcommand{\arraystretch}{1.3}  
\begin{tabular}{cccccc}
 $ v_{BL}$ (GeV) & $\lambda_S$ & $\alpha$ &
 $\beta/H_*$  & $c_s^2$ &
 $T_*$ (GeV)   \\
\hline 
 $10^{7}$ & $2.06\times10^{-4}$ & 8.40 & 147 & 0.3330 &   175159 \\ \hline
 $10^{8}$  &   $2.065\times 10^{-4}$ & 7.33 &   438 & 0.3330 &  1814637 \\ \hline
 $3\times 10^6$  & $2.07\times 10^{-4}$  &  5.49 &   266 & 0.3331 &  58673 \\ \hline
 $6.46\times 10^6$ & $2.05\times 10^{-4}$ &   9.90 &  110 & 0.3330  &  107448
\\
\end{tabular}
\end{ruledtabular}
\caption{
\label{benchmarkpoints_FOPT_lam_1_new} Another set of benchmark values of the FOPT-parameters in the Dirac scenario (first two rows) and in the pseudo-Dirac scenario (last two rows) for the same $m_{Z'}/v_{BL} =0.32$, with different $\lambda_S$. The corresponding GW-spectra is shown in Fig.~\ref{fig:m_plot_new}.
}
\end{table}


\bibliographystyle{utphys}
\bibliography{reference}
\end{document}